\documentclass[prl,twocolumn,floatfix]{revtex4}
\usepackage{graphicx}

\newlength{\onecolfig}
\newlength{\twocolfig}
\newcommand{\ion}[2]{\mbox{$^{#2}$#1$^+$}}
\newcommand{\Ca}[1]{\ion{Ca}{#1}}

\newcommand{\hfslev}[3]{\mbox{#1$^{\mbox{\tiny$#3$}}_{\mbox{\tiny$#2$}}$}}

\newcommand{\unit}[1]{\,\mbox{#1}}
\newcommand{\Hz}{\unit{Hz}}
\newcommand{\kHz}{\unit{kHz}}

\newcommand{\GHz}{\unit{GHz}}

\newcommand{\mV}{\unit{mV}}

\newcommand{\mm}{\unit{mm}}

\newcommand{\nm}{\unit{nm}}

\newcommand{\s}{\unit{s}}
\newcommand{\persec}{\unit{s$^{-1}$}}
\newcommand{\ms}{\unit{ms}}
\newcommand{\us}{\unit{$\mu$s}}

\newcommand{\G}{\unit{G}}
\newcommand{\mG}{\unit{mG}}

\newcommand{\etal}{{\em et al.}}

\newcommand{\ish}{\mbox{$\sim$}\,}
\newcommand{\ltish}{\protect\raisebox{-0.4ex}{$\,\stackrel{<}{\scriptstyle\sim}\,$}}
\newcommand{\gtish}{\protect\raisebox{-0.4ex}{$\,\stackrel{>}{\scriptstyle\sim}\,$}}

\newcommand{\ket}[1]{\mbox{$\left| #1 \right>$}}

\newcommand{\wee}[2]{\mbox{$\frac{#1}{#2}$}}
\newcommand{\sub}[1]{\mbox{$_{\mbox{\tiny #1}}$}}
\newcommand{\diff}[1]{\mbox{\/d$#1$}}

\begin{document}
\bibliographystyle{apsrev}

\title{A long-lived memory qubit on a low-decoherence quantum bus}

\author{D. M. Lucas,  B. C. Keitch, J. P. Home, G. Imreh, M. J. McDonnell, D. N. Stacey, D. J. Szwer and A. M. Steane}
\affiliation{Department of Physics, University of Oxford, Clarendon Laboratory, Parks Road, Oxford OX1 3PU, U.K.}

\date{\today}
\begin{abstract}
We demonstrate long-lived coherence in internal hyperfine states of a single \Ca{43} trapped-ion qubit $[T_2=1.2(2)\s]$, and in 
external motional states of a single \Ca{40} trapped-ion qubit $[T_2'=0.18(4)\s]$, in the same apparatus. The motional decoherence 
rate is consistent with the heating rate, which was measured to be 3(1) quanta/sec. Long coherence times in the external motional 
states are essential for performing high-fidelity quantum logic gates between trapped-ion qubits. The internal-state $T_2$ time 
that we observe in \Ca{43}, which has not previously been used as a trapped-ion qubit, is about one thousand times longer than 
that of physical qubits based on \Ca{40} ions. Using a single spin-echo pulse to ``re-phase'' the internal state, we can detect no 
decoherence after 1\s, implying an effective coherence time $T_2^{\mbox{\tiny SE}} \gtish 45\s$. This compares with timescales in 
this trap for single-qubit operations of \ish 1\us, and for two-qubit operations of \ish 10\us. 
\end{abstract}


\maketitle


Trapped atomic ions appear to offer one of the most promising systems for the realization of quantum information processors. The 
fundamental prerequisites of qubit state preparation, two-qubit entangling gates and state readout have all been achieved with a 
fidelity of 97\% or higher~\cite{03:Leibfried, 06:Riebe, 05:Leibfried, 05b:Haffner}, and these operations have been made 
qubit-specific by means of either tightly-focused laser beams~\cite{04:Riebe} or shuttling ions between multiple trapping 
regions~\cite{04:Chiaverini}. Detailed proposals exist for scaling up these systems to several tens of qubits and 
beyond~\cite{98:Wineland,07:Steane}.

One of the challenges facing any proposed quantum computing implementation is that of obtaining coherence times sufficiently long 
(compared with qubit operation times) that the error thresholds necessary for fault-tolerant computation can be 
attained~\cite{07:Steane}. Coherence must be maintained both in the internal qubit states and during whatever process is used to 
perform quantum logic gates between different qubits. In ion trap approaches,
the internal states are either ground level hyperfine or Zeeman states, or long-lived metastable states; with recent 
exceptions~\cite{05:Langer, 05:Haljan} the states used have been sensitive in first order to magnetic fields and limited to a 
coherence time $T_2\ltish 1\ms$ by magnetic field noise. Pairs of ions have been used to encode a single logical qubit in a 
decoherence-free sub-space which is immune to magnetic field noise~\cite{05:Langer,05:Haffner}, to achieve qubit coherence times 
of \ish 10\s; however this involves an encoding/decoding overhead and it is clearly advantageous to start with physical qubits 
which are themselves stable. All demonstrations of entangling gates in ion traps have relied on the shared modes of external 
motion to couple the trapped-ion qubits and are thus also sensitive to decoherence in the motional states; this decoherence arises 
mostly from noise in the electric fields used to confine the ions. Motional coherence times have been directly measured in a 
\ion{Be}{9} trap~\cite{00b:Turchette} ($T_2'\approx 0.2\ms$) and a \ion{Ca}{40} trap~\cite{03:SchmidtKaler} ($T_2'\approx 
100\ms$).

In this Letter we demonstrate long-lived internal qubit coherence in a trap which also shows low decoherence of
the motional states needed for quantum logic gates. The long-lived internal coherence is achieved by using ground
hyperfine ``clock'' states of \Ca{43} which are nearly independent of magnetic field to first order. This ion has not
previously been used as a qubit but has a number of advantages over other candidate species. The required laser
wavelengths are convenient and available from solid-state diode lasers without frequency-doubling. Recent
calculations~\cite{06:Ozeri} show that the fundamental photon-scattering infidelity for gates based on Raman
transitions can meet fault-tolerant requirements ($\ish 10^{-4}$) at attainable laser powers~\cite{07:Keitch}. The
existence of metastable D states allows very high fidelity ($>99.9\%$) qubit readout by the technique of electron
shelving~\cite{07:Szwer}. To measure the motional coherence time in this trap we use a \Ca{40} ion-qubit: the even
isotope is used for technical convenience and we would expect similar results for \Ca{43}.


We denote the \Ca{43} qubit states by $\ket{\uparrow}=\hfslev{S}{1/2}{3,0}$ and $\ket{\downarrow}=\hfslev{S}{1/2}{4,0}$ (where the 
superscript indicates the angular momentum quantum numbers $F,M_F$). A single \Ca{43} ion is loaded into the ion trap from a 
natural abundance (0.14\%) source by isotope-selective photoionization, where it may be held for several days (see \cite{04:Lucas} 
for apparatus details and \Ca{43} level diagram). The ion is Doppler-cooled on the 
$\hfslev{S}{1/2}{4}\leftrightarrow\hfslev{P}{1/2}{4}$ and $\hfslev{S}{1/2}{3}\leftrightarrow\hfslev{P}{1/2}{4}$ transitions which 
are driven respectively by a 397\nm\ laser and a 3.220\GHz\ laser sideband (provided by an electro-optic modulator, EOM). A 
single-frequency 866\nm\ laser repumps on the $\hfslev{D}{3/2}{}\leftrightarrow\hfslev{P}{1/2}{}$ transition. We can prepare the 
\hfslev{S}{1/2}{4,+4} state with \gtish 99\% efficiency by optical pumping with a circularly-polarized 397\nm\ beam. However, the 
required beam direction for efficient optical pumping to one of the $M_F=0$ qubit states is not available in this apparatus; 
instead we simply switch off the EOM, which prepares the \ket{\uparrow} state with \ish 14\% efficiency (and we ignore those 
experiments in which some other \hfslev{S}{1/2}{3,M_F} state was prepared). The 
$\hfslev{S}{1/2}{4,0}\leftrightarrow\hfslev{S}{1/2}{3,0}$ qubit transition is directly driven by 3.226\GHz\ microwaves applied to 
a spare pair of trap electrodes. The microwaves can also be used to probe the magnetic-field-sensitive 
$\hfslev{S}{1/2}{4,+4}\leftrightarrow\hfslev{S}{1/2}{3,+3}$ transition in order to measure the magnetic field. After applying the 
microwaves, we read out the state of the qubit by selectively shelving the \ket{\downarrow} state in the metastable 
\hfslev{D}{5/2}{} level with a short pulse of circularly-polarized 393\nm\ light tuned to the 
$\hfslev{S}{1/2}{4}\leftrightarrow\hfslev{P}{3/2}{5}$ transition (the \ket{\uparrow} state is shelved with probability $<0.2\%$). 
The Doppler-cooling lasers are then re-applied and the presence or absence of fluorescence indicates whether the ion was shelved. 
The single-shot fidelity of this readout process is \ish 95\%, limited mainly by decay on the 
$\hfslev{P}{3/2}{}\rightarrow\hfslev{D}{3/2}{}$ transition. Finally, the ion is de-shelved by driving 
$\hfslev{D}{5/2}{}\leftrightarrow\hfslev{P}{3/2}{}$ with an 854\nm\ laser.

\begin{figure}
\includegraphics[width=\onecolfig]{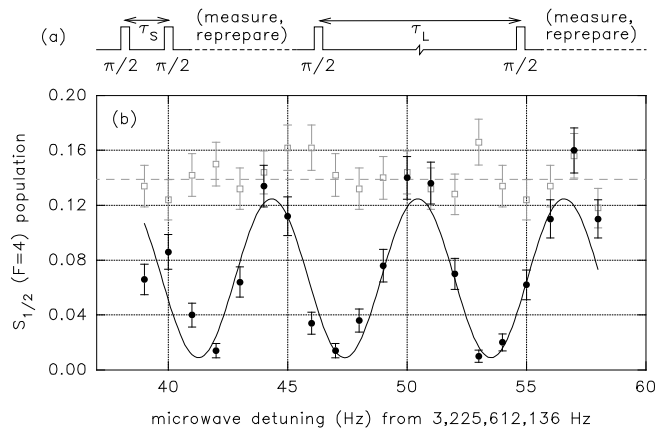}
\caption{%
Internal state coherence measurement using \Ca{43}.
(a) Simplified pulse sequence, showing the control experiment with short Ramsey delay $\tau_S$ followed by the long Ramsey 
experiment with delay $\tau_L$. The qubit is measured and reprepared after each Ramsey experiment.
(b) Typical Ramsey fringes, with $\tau_L=200\ms$. The filled symbols show the mean \hfslev{S}{1/2}{4} population measured at the 
end of the long Ramsey sequence, as a function of detuning from the centre of the 
$\hfslev{S}{1/2}{4,0}\leftrightarrow\hfslev{S}{1/2}{3,0}$ qubit transition. The empty symbols show the \hfslev{S}{1/2}{4} 
population after the short Ramsey sequence, with the mean level indicated by the horizontal dashed line. Binomial error bars are 
plotted. The curve is a sinusoidal fit to the data, which has peak-to-peak amplitude 0.85(10) relative to the control level. The 
total run duration was 37\unit{min}.
}
\label{F:ca43ramsey}
\end{figure}

To determine the coherence time of the qubit states, we measure the amplitude of the fringes in Ramsey experiments on the 
$\ket{\downarrow}\leftrightarrow\ket{\uparrow}$ qubit transition as a function of the delay time $\tau_L$ (up to 300\ms) between 
the two Ramsey $\pi/2$ pulses, each of duration 35\us. The pulse sequence, shown in fig.~\ref{F:ca43ramsey}(a), actually consists 
of two pairs of Ramsey pulses, the first pair with a short (0.145\ms) delay $\tau_S\ll\tau_L$ to act as a control. Typically 500 
such sequences were performed at a particular microwave frequency to allow the mean \hfslev{S}{1/2}{4} population to be 
determined. The frequency was then stepped so that in each experimental run data were obtained over a typical frequency range 
$(\pm 1/\tau_L)$ close to resonance so as to produce standard Ramsey fringes as in fig.~\ref{F:ca43ramsey}(b). The control 
experiment gives a check for drift (for example, in readout fidelity): $\tau_S$ is so short that the pulses effectively constitute 
a single $\pi$ pulse and the readout gives the ``short-time'' amplitude to which the fringes from the long experiment can be 
normalized.  Since the control and test experiments are interleaved in time there is no significant delay between them. The 
control experiments showed no significant drift, even during the longest experiments (total run duration 54\unit{min}). In some 
runs, the phase of the second $\pi/2$ pulse was swept relative to that of the first; in this case both the control and test 
experiments give Ramsey fringes. For the majority of runs the microwave synthesizer was locked to a rubidium-referenced quartz 
crystal, with stability (Allan deviation) below $2\times 10^{-11}$ for times $1\s < t < 10^6\s$ and thus negligible drift over the 
timescales of these experiments.

Ramsey fringes from an experiment with $\tau_L=200\ms$ are shown in fig.~\ref{F:ca43ramsey}(b), together with the reference level 
from the control experiment. The fit gives peak-to-peak amplitude 0.85(10) relative to the control, and period 6.1(1)\Hz\ which 
differs significantly from the expected period of $1/\tau_L=5.0\Hz$. At the working magnetic field of 1.78\G, the residual 
magnetic field sensitivity of the qubit transition due to the second-order Zeeman shift is 4.33\unit{Hz/mG}. A linear field drift 
of $-1.3\unit{mG/hr}$ over the 37\unit{min} run duration would account for the incorrect fringe period, but would reduce the 
fringe amplitude by a negligible amount. The reduction in amplitude is however consistent with faster field fluctuations at the 
level of $\ish 0.25\mG$ over timescales of a few minutes. Both fluctuations and long-term drift of the magnetic field at these 
levels were observed in the trap in independent measurements on the field-dependent transition 
$\hfslev{S}{1/2}{4,+4}\leftrightarrow\hfslev{S}{1/2}{3,+3}$ and are typical for an apparatus where there is no magnetic shielding 
or active field stabilization.

A Ramsey spectrum such as fig.~\ref{F:ca43ramsey}(b), together with a measurement of the magnetic field obtained from the 
$\hfslev{S}{1/2}{4,+4}\leftrightarrow\hfslev{S}{1/2}{3,+3}$ transition frequency, can be used to determine the \Ca{43} ground 
state hyperfine splitting by using the Breit-Rabi formula to extrapolate to zero field. Including uncertainty in the rubidium 
clock reference, we find a value 3225608288(3)\Hz, in good agreement with that of 3225608286.4(3)\Hz\ reported in~\cite{94:Arbes}. 
The level of agreement indicates that other systematic effects which might perturb the qubit energy splitting and lead to 
decoherence are likely to be \ltish 3\Hz.

\begin{figure}
\includegraphics[width=\onecolfig]{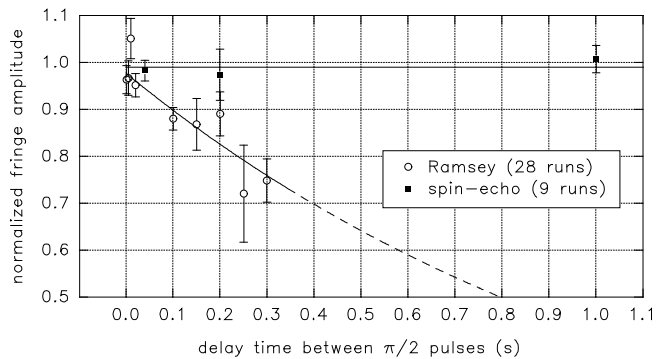}
\caption{%
Amplitude of internal state Ramsey and spin-echo interference fringes as a function of delay between the two $\pi/2$ pulses, 
normalized using the control experiments. For clarity, different experimental runs with the same delay have been combined into a 
single data point: there were 28 Ramsey runs (of which 6 were at 0.3\s) and 9 spin-echo runs (of which 3 were at 1\s). The Ramsey 
experiments were taken in 5 separate series, each series on a different day. The exponential decay curve is a weighted fit to the 
(uncombined) data; it has time constant $T_2=1.2(2)\s$, intercept 0.98(2) and reduced $\chi^2_{\nu}=0.83$. The spin-echo runs were 
taken over two days; their amplitude shows no detectable decay and has a weighted mean of 0.99(1).
}
\label{F:ca43T2}
\end{figure}

Results from five series of Ramsey experiments (28 runs) are shown combined in fig.~\ref{F:ca43T2}.
In each series the long Ramsey delay $\tau_L$ was varied between runs; each series was taken on a different day. From each series 
we can extract a value for the coherence time $T_2$ by fitting an exponential decay to the measured fringe amplitude. Values of 
$T_2$ thus extracted do not show significant differences, and we combine the entire dataset by fitting $\alpha\exp(-\tau_L/T_2)$ 
to the normalized fringe amplitude. This gives $T_2=1.2(2)\s$ for the internal state coherence time.

\begin{figure}
\includegraphics[width=\onecolfig]{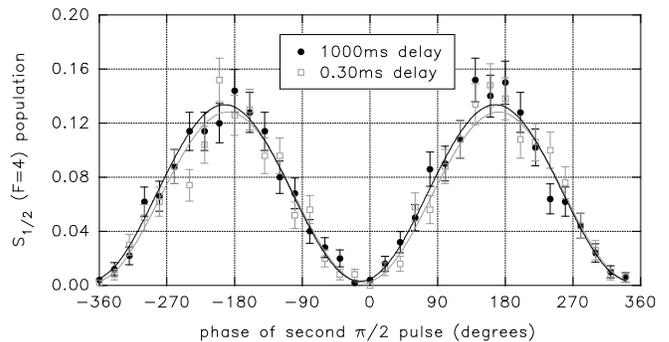}
\caption{%
Internal state spin-echo experiment, with a delay of 1000\ms\ between the two $\pi/2$ pulses. The filled symbols show the mean 
\hfslev{S}{1/2}{4} population, measured at the end of the 1000\ms\ spin-echo sequence, as a function of the phase of the second 
$\pi/2$ pulse. The empty symbols show the mean \hfslev{S}{1/2}{4} population at the end of the interleaved control experiment (a 
0.30\ms\ spin-echo sequence, where the phase was also scanned). Both data sets are fitted with sinusoidal functions and the fitted 
amplitude of the 1000\ms\ fringes is 1.03(4) relative to that of the control. The phase offset is arbitrary. The total duration of 
the run was 5.2 hours.
}
\label{F:spinecho}
\end{figure}

The reduction in fringe amplitude observed at long $\tau_L$ may be due to magnetic field fluctuations, as described above, which 
dephase the qubit, or to other processes such as photon scattering by imperfectly extinguished laser beams, which project the 
qubit. The presence of magnetic field fluctuations alone does not constitute decoherence of the qubit, in the sense of an 
entanglement with the environment which would require the techniques of quantum error-correction to recover. Slow drift, in either 
the qubit precession frequency (caused by the magnetic field drift) or the microwave oscillator, can be corrected for much more 
simply with the well-known spin-echo technique: a $\pi$ pulse inserted in the centre of the Ramsey gap has the effect of 
``re-phasing'' the qubit superposition provided that fluctuations are slow compared with the total length of the spin-echo 
sequence. When this was done, we could observe no measurable decoherence of the qubit (fig.~\ref{F:ca43T2}). The longest spin-echo 
sequence attempted was 1000\ms\ (fig.~\ref{F:spinecho}) and from the three experimental runs with this Ramsey delay we place a 
($1\sigma$) lower limit on the coherence of 98\% at 1\s. This indicates that amplitude and phase decoherence of the qubit (such as 
might be caused by residual photon scattering, or collisions with background gas atoms) are negligible over this timescale. If the 
noise model for any remaining decoherence processes is taken to be a simple exponential, as in the Ramsey experiments, this limit 
would imply an effective coherence time $T_2^{\mbox{\tiny SE}}>45\s$. Although the fringe amplitude showed no decay out to 1\s, a 
small increase in the fringe baseline was detectable, at the level of 0.03(2)\persec\ relative to the amplitude; this baseline is 
more sensitive to certain decoherence processes as all seven \hfslev{S}{1/2}{3,M_F} states contribute to the signal, not just 
\hfslev{S}{1/2}{3,0}. Any mechanism (for example, photon scattering) which redistributes population from \hfslev{S}{1/2}{3} to 
\hfslev{S}{1/2}{4} will increase the baseline. Alternatively, occasional heating of the ion during the long spin-echo delay could 
cause fluorescence to be lost which would be registered as the ion being shelved. If, as a worst case, we interpret this as 
entirely due to a real decoherence process it would indicate qubit decoherence at the level of $\ish 0.5\%$ at 1\s, or an 
effective coherence time of between 2 and 10 minutes.

To measure the motional coherence time in the same ion trap, we used the $\ket{\uparrow}=\hfslev{S}{1/2}{+1/2}$ and 
$\ket{\downarrow}=\hfslev{S}{1/2}{-1/2}$ Zeeman states of the ground level of a single \ion{Ca}{40} ion (where the superscript now 
refers to the quantum number $M_J$). The spin state detection is described in~\cite{04:McDonnell}. Ramsey experiments were 
performed in the basis $(\ket{\downarrow,n=0}, \ket{\downarrow,n=1})$ where $n$ represents the vibrational quantum number 
associated with the axial motion of the ion. The axial trap frequency was 810\kHz. The Raman carrier transition 
$\ket{\downarrow}\leftrightarrow\ket{\uparrow}$ and motional-sideband transitions, for example 
$\ket{\uparrow,n=0}\leftrightarrow\ket{\downarrow,n=1}$, are driven coherently by a pair of Raman laser beams at 397\nm, detuned 
by +30\GHz\ from the $\hfslev{S}{1/2}{}\leftrightarrow\hfslev{P}{1/2}{}$ transition.

\begin{figure}
\includegraphics[width=\onecolfig]{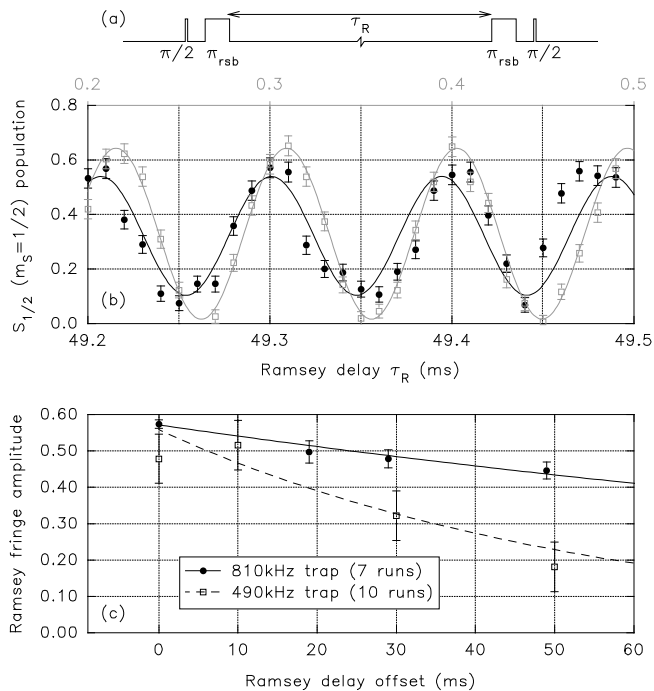}
\caption{%
Motional state coherence experiment using \Ca{40}.
(a) Pulse sequence for motional Ramsey experiment ($\pi\sub{rsb}$ = red-sideband $\pi$ pulse).
(b) Typical Ramsey fringes where the detuning from the 810\kHz\ motional resonance was $\delta_M=10.7(2)\kHz$. The empty symbols 
(upper abscissa) are for a run with Ramsey delay offset $\tau_M=0.2\ms$, the filled symbols (lower abscissa) are for a run with 
$\tau_M=49.2\ms$. The mean \hfslev{S}{1/2}{+1/2} population has been corrected for readout error.
(c) Ramsey fringe amplitude as a function of the Ramsey delay offset $\tau_M$. Different runs with the same $\tau_M$ have been 
combined into a single data point for clarity. Two datasets are shown, for 810\kHz\ and 490\kHz\ axial trap frequencies. The 
exponential decay fits to the (uncombined) data have time constants $T_2'=182(36)\ms$ and $T_2'=56(18)\ms$ respectively.
}
\label{F:motion}
\end{figure}

The ion is first cooled close to the $n=0$ motional ground state (typically $\bar{n}<0.1$) by three stages of laser cooling: 
Doppler cooling for 2\ms, continuous Raman sideband cooling for 4\ms\ and ten pulses of pulsed sideband Raman cooling taking 
0.5\ms. A Raman carrier $\pi/2$ pulse (duration 1.8\us) prepares the superposition $\wee{1}{\sqrt{2}} 
(\ket{\downarrow,n=0}+\ket{\uparrow,n=0})$, then a red-sideband $\pi$ pulse (20\us) immediately maps this to the state 
$\wee{1}{\sqrt{2}} (\ket{\downarrow,n=0}+\ket{\downarrow,n=1})$. After a Ramsey delay $\tau_R$ the same $\pi$ and $\pi/2$ pulses 
are applied in the reverse order, fig.~\ref{F:motion}(a). In these experiments, the frequency of the red-sideband pulse was held 
constant, but detuned from the motional trap frequency by $\delta_M \approx 10\kHz$, while the delay $\tau_R$ was swept over an 
interval $(\tau_M,\tau_M + 3/\delta_M)$ sufficient to produce three Ramsey fringes. Typical fringes are shown in 
fig.~\ref{F:motion}(b). The motional coherence time was determined by increasing the delay offset $\tau_M$ and measuring the 
decrease in fringe amplitude, fig.~\ref{F:motion}(c). We find a decay constant $T_2'=182(36)\ms$. Similar measurements were made 
at an axial trap frequency of 490\kHz, and gave a motional coherence time of $T_2'=56(18)\ms$.

The motional coherence time in the 810\kHz\ trap is longer than that reported for other ion traps, even before scaling for the 
relatively low trap frequency, but is comparable to the time scale at which motional heating takes place in this trap. In 
independent experiments, we have measured heating rates as low as $\diff{\bar{n}}/\diff{t}=3(1)\persec$ at a trap frequency of 
810\kHz~\cite{Th:Home}; this is also the lowest reported heating rate for room-temperature trap electrodes, but is consistent with 
that expected given the relatively large ion-electrode distance $\rho=1.2\mm$ and the empirically measured $\propto 1/\rho^4$ 
scaling~\cite{07:Epstein}. The heating rate is most sensitive to electric field noise in the vicinity of the trap frequency, 
whereas motional decoherence can also be caused by, for example, slow fluctuations in the electric field. The measured $T_2'$ in 
both the trap strengths investigated here is consistent with voltage fluctuations on the trap end-cap electrodes at the level of 
\ish 3\mV, which is well within the specified limits of the voltage supply. Slow fluctuations could be compensated for by a 
spin-echo technique analogous to that used for the internal state measurements above, but this is unlikely to be useful during a 
quantum logic gate---the very operation where motional decoherence is a concern.


In conclusion, we have demonstrated a long-lived memory qubit, coupled to a motional quantum ``bus'' with extremely low 
decoherence. Using a simple spin-echo pulse, the decoherence of the memory qubit was suppressed below measurable levels, 
indicating an effective coherence time at least as long as that measured for any single physical qubit. The microwave spin-echo 
technique is attractive as it can be applied simultaneously to all qubits in a quantum processor with high precision and without 
the photon-scattering that accompanies laser techniques. The ratio between decoherence time and quantum logic operation time is 
$\ish 10^6$.


This work was supported by the Disruptive Technology Office (contract W911NF-05-1-0297), the EPSRC (QIP IRC), the European Union 
(CONQUEST and SCALA networks) and the Royal Society. We thank J.~Sherman and S.~Webster for helpful comments on the manuscript.

\end{document}